# AI2T: Building Trustable AI Tutors by Interactively Teaching a Self-Aware Learning Agent

DANIEL WEITEKAMP, ERIK HARPSTEAD, and KENNETH KOEDINGER, Carnegie Mellon University, USA

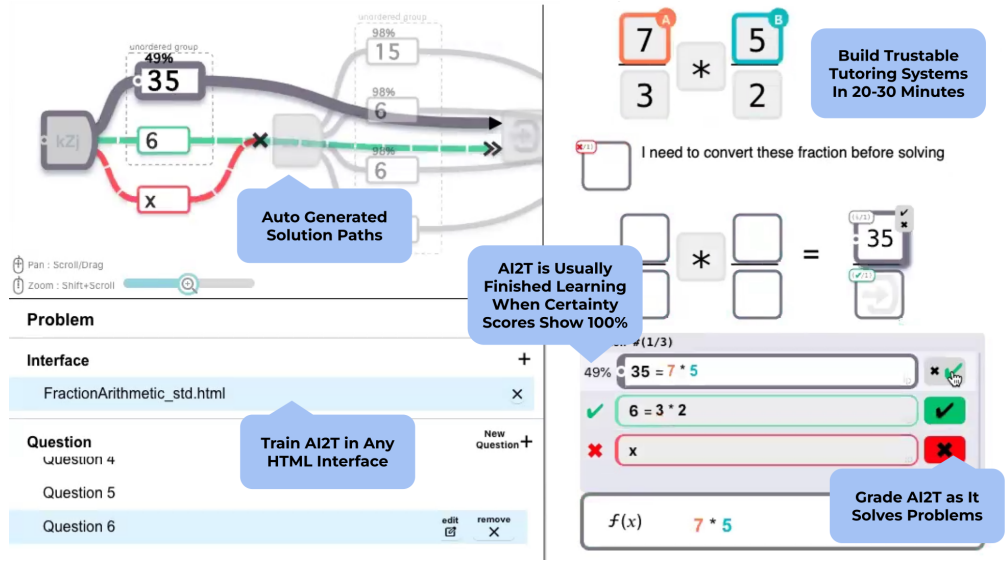

Fig. 1. Authors tutor AI2T in HTML interfaces and grade its step-by-step solutions. Certainty scores help authors determine when AI2T has learned robust tutoring programs.

AI2T is an interactively teachable AI for authoring intelligent tutoring systems (ITSs). Authors tutor AI2T by providing a few step-by-step solutions and then grading AI2T's own problem-solving attempts. From just 20-30 minutes of interactive training, AI2T can induce robust rules for step-by-step solution tracking (i.e., model-tracing). As AI2T learns it can accurately estimate its certainty of performing correctly on unseen problem steps using STAND: a self-aware precondition learning algorithm that outperforms state-of-the-art methods like XGBoost. Our user study shows that authors can use STAND's certainty heuristic to estimate when AI2T has been trained on enough diverse problems to induce correct and complete model-tracing programs. AI2T-induced programs are more reliable than hallucination-prone LLMs and prior authoring-by-tutoring approaches. With its self-aware induction of hierarchical rules, AI2T offers a path toward trustable data-efficient authoring-by-tutoring for complex ITSs that normally require as many as 200-300 hours of programming per hour of instruction.

Authors' Contact Information: Daniel Weitekamp, weitekamp@cmu.edu; Erik Harpstead, harpstead@cmu.edu; Kenneth Koedinger, koedinger@cmu.edu, Carnegie Mellon University, Pittsburgh, Pennsylvania, USA.

## 1 Introduction

A dominant theme of the last decade of AI research has been to use big-data machine learning to replicate patterns of behavior distributed across large datasets or explore uses for pretrained models that have been produced in a data-driven manner. By contrast, AI2T (pronounced *A.I. tutee*) is an AI agent that can be **T**aught interactively, in a process that induces **T**rustable well-defined educational programs. These are the two **T**s of AI2T. A central issue with trusting many AI capabilities trained with machine learning is that their learned behaviors are rarely errorless or internally inspectable. Popular methods like deep-learning [31] fit high-dimensional neural networks with inscrutable 'blackbox' weights that can produce very flexible, but often inconsistent behaviors. AI2T avoids these particular issues of trust by inducing well-defined programs expressed as hierarchical task networks (HTNs) [22]. AI2T induces executable symbolic generalizations that are consistent with the author's training.

One way to trust machine-synthesized programs is to inspect the induced programs directly. However, successful code checking requires some programming proficiency and a familiarity with the representation language targeted by the method of program synthesis. AI2T attempts to guide users toward building trustable programs in a different manner. It employs a machine learning method called STAND that can learn efficiently from very little data, yet still accurately estimate its prediction certainty on unseen examples [57]. When STAND's certainty scores tend to increase its true holdout set performance tends to increase. This allows AI2T to be self-aware of its learning, and users can use STAND's certainty scores as an indicator of when AI2T's induced program is complete and trustable. We show in simulation that changes in STAND's certainty estimates more accurately reflect actual changes in holdout set performance than competing predictions from more conventional (and relatively data-efficient) ensemble methods like random forests and XGBoost, which happen to only be as good as chance in this regard. Additionally, we show in a user study that users can successfully use STAND's certainty estimates, as a heuristic for deciding when AI2T has been trained on sufficient practice problems. STAND enables AI2T to be essentially self-aware of its learning progress. This helps users estimate when they have provided AI2T with enough training examples to induce programs with 100% correct behavior.

The core features of AI2T span several machine learning paradigms including machine teaching [60], interactive machine learning, [15], and interactive task learning (ITL) [28]. At its core AI2T builds upon the programming-by-demonstration (PBD) paradigm, where non-programmers demonstrate program behavior instead of writing code in a programming language [13]. Many demonstrations of PBD have shown robust performance in automating simple single action or sequential behaviors [18, 29, 34] that require minimal generalizations from users' demonstrations. However, the challenges of PBD multiply when applied to building larger multi-faceted applications [40, 43]. AI2T can very robustly induce the core behaviors of complex applications known as Intelligent Tutoring Systems (ITS)—educational technologies known for their comprehensive adaptive student support features. Authors train AI2T with a set of interactions that go beyond PBD. These interactions are better described as authoring-by-tutoring [39]—authoring with AI2T involves both demonstrating solutions and interactively checking AI2T's behavior as it attempts to solve problems on its own. In this relationship, the author is the tutor, and AI2T is the tutee. Therefore AI2T is best described by Laird et. al.'s vision of interactive task learning (ITL) [28], the idea of AI systems that can be taught complex and robust new capabilities using a variety of natural interactions that are intuitive to non-programmers.

### 1.1 Intelligent Tutoring Systems: The Original AI Tutors

Long before recent hype about tutoring with generative AI, hand-programmed expert-system-like AI were used to build Intelligent Tutoring Systems (ITS). Decades of learning science research has honed a set of best practices for designing these conventional ITSs [17, 53]. ITSs have been shown to be more effective than traditional classroom instruction alone [53], and in some cases more effective than traditional human-to-human tutoring [27]. The key to historical ITS successes has been instruction designed around learning-by-doing exercises where the ITS provides step-by-step cognitive assistance that adaptively supports students' directed practice [11, 26].

For instance, model-tracing tutors track students' step-by-step solutions to determine in what ways their current knowledge aligns with or diverges from a model of expert knowledge [24, 44]. Model-tracing enables ITSs to directly track and adapt instruction to student's misconceptions and unmastered knowledge as it tracks their progress through active step-by-step practice. When AI2T learns ITS behavior it induces rule-based knowledge structures sufficient for executing this historically difficult-to-build class of ITS behaviors. Model-tracing ITSs have been estimated to require as many as 200-300 developer hours per hour of instruction [1]. In the two domains we use to evaluate this work, AI2T cuts the most difficult programming elements of authoring down to about 20-30 minutes of effort. Authoring with AI2T is similar to tutoring a human, and involves mostly just solving problems and grading AI2T as it solves problems.

AI2T can induce correct and complete model-tracing behavior in the sense that it induces rules that permit only the correct next actions in step-by-step problem solving, and no incorrect next actions. This induced behavior also permits the generation of "bottom-out" hints: correct next actions for problem steps that are requested by students as a last resort when they are stuck [19]. Typically other features of ITSs (which we do not focus on in this work) like requestable conceptual hints, automatic feedback messages, and knowledge-tracing (i.e. tracking students' mastery of particular knowledge) [12] are built on top of the production rules of a model-tracer.

Readers may fairly wonder why we have not taken an LLM-based approach. Large Language Models (LLMs) have many exciting applications in education but have yet to demonstrate behaviors similar to model-tracing tutors. Insofar as out-of-the-box LLM chatbots 'tutor', they typically default to providing full step-by-step explanations similar to textbook worked examples [45]—an impressive feature no doubt, but one that is prone to abuse, and inconsistent with historically successful ITS designs that focus on learning-by-doing exercises. Even if an LLM is prompted or fine-tuned to evaluate intermediate steps of a student solution in-progress (i.e., perform model-tracing), there is no guarantee that it will provide consistently accurate evaluations and, indeed, its underlying statistical inference approach makes 100% accuracy highly unlikely. Moreover, LLMs often produce "hallucinations": responses that often sound plausible but provide information that is misleading, logically inconsistent, or entirely fabricated. In an educational setting, even a small rate of hallucinated incorrect instruction may do more harm than good—plausible but incorrect responses are a likely recipe for producing student misconceptions. It is hard to beat the precision of a robust well-defined program. AI2T innovates toward authoring these trustable tutoring programs without writing code.

## 1.2 Traditional ITS Authoring

Several tools like CTAT example-tracing, OATutor [47], and others [3, 21, 46] offer approaches that are faster than programming-based authoring and accessible to non-programmers. Yet these methods place considerable limits on ITS control structures. OATutor supports strictly sequential "tutoring pathways" [47], and CTAT example-tracing supports graphs of states and actions that can diverge, re-converge, and manifest unordered groups—essentially limiting them to a slightly larger class of finite state machines-like control structures [1]. Both CTAT example-tracing and OATutor enable means of mass-producing problems within these fixed control structures via a template-filling approach, where special variable strings are replaced by problem-specific content detailed in spreadsheets. In practice, this method of mass-producing problems still requires some programming effort, for instance, by writing programs to fill in the content of each step in a spreadsheet formula language [1, 47]. By contrast, authoring with AI2T requires no programming (or program checking), but can produce behaviors that are typically only implementable by programming.

The domains in which these programmed rule-based model-tracing ITSs are particularly useful include complex procedural skills that involve context-specific decision making, and multi-step domains that should permit some solution flexibility that is impractical to capture in fixed sequential or graph-based control structures. STEM-based procedural tasks are a major category of domains where these features are desirable, yet any domain that teaches complex procedural skills is a good candidate for the kinds of model-tracing behaviors that AI2T can author.

### 1.3  Issues with prior Authoring-by-Tutoring Approaches

Prior authoring-by-tutoring approaches like those prototyped with SimStudent [39] and the Apprentice Learner (AL) [37, 54], have demonstrated efficiency benefits over conventional authoring tools. However, in studies of these prior approaches, untrained participants [54], and in some cases also the creators of those approaches [37, 39, 55], failed to produce ITS behavior that was 100% model-tracing complete. *Model-tracing completeness* is the proportion of reachable problem states across a large holdout set of problems where an agent or ITS permits every correct next action (defined by a ground-truth ITS) and no incorrect actions. Weitekamp et. al reported that their authoring-by-tutoring approach with AL fell short of 100% model-tracing completeness both because of limitations in its interaction design and the learning mechanisms of their agent [54]. For instance, they report interaction design issues related to participants locating and fixing mistakes, navigating and providing feedback over diverging solution paths, and estimating when training is complete. AI2T builds on the interaction designs and machine-learning approaches of prior work to resolve many of these issues.

## 2  Two Evaluated Domains: And Beyond

In this work, we evaluate AI2T in simulation experiments and in two user studies. For both of these evaluation methods, we restrict ourselves to two domains: multicolumn addition and fraction arithmetic. Both domains require some contextual decision making and permit some solution flexibility (in terms of step order). Expressed as heirarchical procedures both domains involve deciding between alternative subprocedures based on the context of the problem. Multicolumn addition in particular is a good example of a domain where an author may resort to programming a model-tracing ITS instead of using conventional graphical authoring tools because the individual steps required to solve different problem instances vary considerably.

In multicolumn addition, students practice the algorithm for summing large numbers together by computing partial sums and carrying their tens digit (if necessary). The ground-truth model-tracing behavior permits add and carry actions to be applied in either order. The main difficulty of this domain in terms of inducing the correct behavior is that a contextual decision must be made about when to carry a 1 or not, and for always adding three numbers instead of two if a 1 was carried from the previous column. To replicate prior work [54] we limit this domain to problem instances that have pairs of 3-digit numbers.

In fraction arithmetic, students must correctly select and apply one of three arithmetic procedures (add, multiply, or convert-then-add). In this domain, fractions are converted by simply multiplying their denominators, and then multiplying crosswise to find the converted numerators. This tutoring system partially scaffolds the process of determining when fractions need to be converted. If conversion is necessary students check a box labeled: "I need to convert these fractions before solving". The ground-truth model-tracing behavior for this domain permits applying the four steps for converting the two numerators and two denominators in any order. The final two steps for computing the combined (added or multiplied) fraction can also be done in either order. In both domains, the student must press a 'done' button as a final action.

This work focuses on evaluating whether AI2T's interaction design and machine learning capabilities overcome issues reported by prior authoring-by-tutoring approaches. We do not evaluate the breadth of what AI2T can author in this work beyond these two domains. Prior work with Sierra [52], SimStudent [39], and the Apprentice Learner [36], which learn in a similar manner as AI2T, have trained agents on dozens of domains including algebra, stoichiometry [33], geometry [8], simple linguistic tasks like Chinese character translation and article selection [35], a few games, and other arithmetic tasks like subtraction [52] and multiplication. Future work may evaluate whether AI2T can succeed where these prior systems have fallen short of inducing correct and complete behavior. In this work, we've focused on just two domains that we could expect untrained participants to remember how to do and author in a 90-minute session.

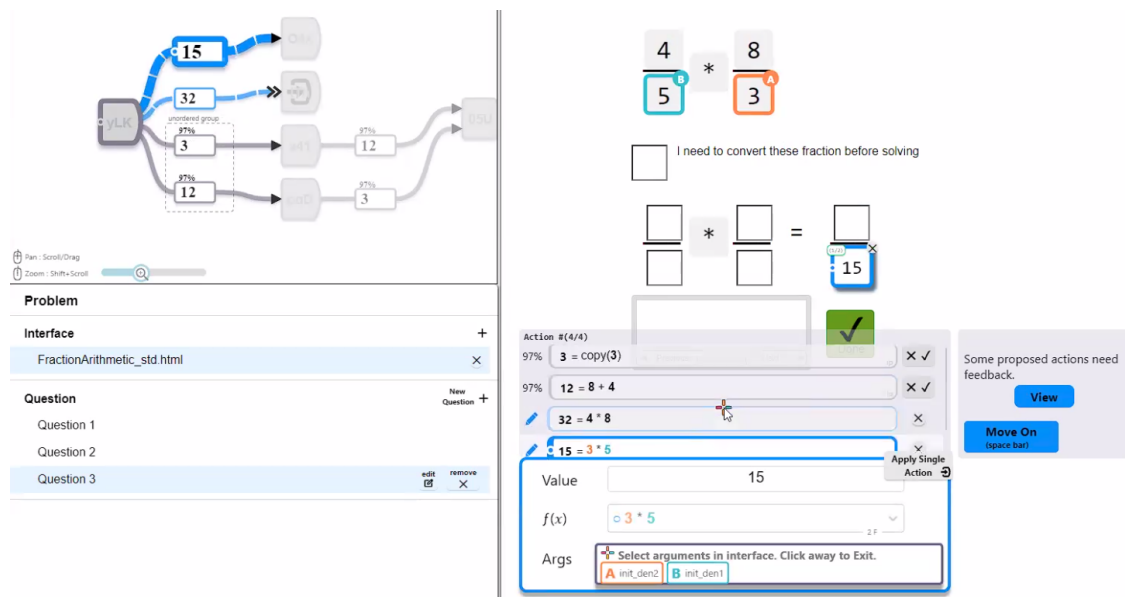

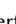

Fig. 2. AI2T's interface. The user has just demonstrated two actions. They show up as blue dashed edges in the behavior graph (top-left) and are indicated by ✏️ in the skill application window (middle-bottom). One of the demonstrations for multiplying the denominators of the expression 4/5 * 8/3 is selected and is previewed as a 15 in the tutor interface overlay. Demonstrations can be removed by clicking the X button in the skill application window or on the tutor interface. Two other actions were proposed by the agent as part of an unordered group (grey edges in dashed box). They both have high certainty scores of 97%, but are incorrect. The user can add correctness feedback by clicking the ✗ or ✓ icons on the toggler in the skill application window.

## 3  AI2T's Interaction Design

Beyond recalling how to solve problems in each domain, AI2T does not require authors to have a particularly specialized skill set. The core interactions for training AI2T are typical tutoring interactions: demonstrating problem solutions, and giving feedback to AI2T on its problem attempts. Beyond this, AI2T's interaction design imposes just a few responsibilities on the author:

(1)  The author needs to provide an interface for their tutoring system.

(2)  The author needs to double-check that AI2T has correctly interpreted their demonstrated actions.

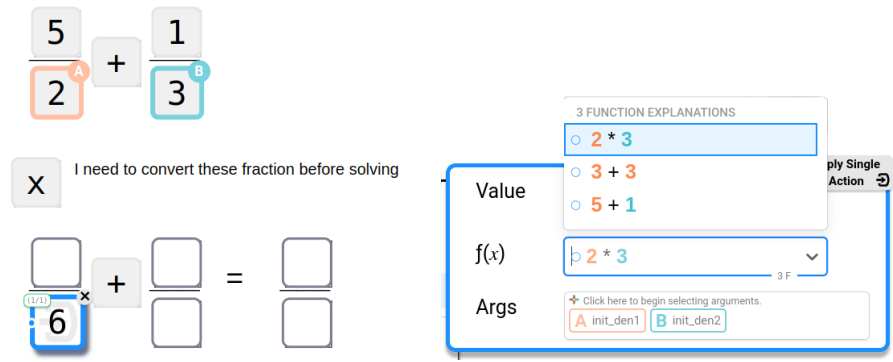

Fig. 3. (a) The user has demonstrated the converted fraction 6. (b) The agent displays several possible explanations for this demonstration in a drop-down. The arguments of the correct explanation are currently highlighted in the tutor interface.

(3) The author needs to provide correctness feedback (a binary yes or no) to all actions that AI2T proposes for each problem state. It is not enough to mark only single solution paths as correct.

(4) The author needs to ensure that AI2T consistently proposes all solution paths that they would want to permit as correct for each training problem.

(5) The author needs to train AI2T on a sufficient number of diverse problems and decide at what point its induced program can be trusted on unseen problems.

AI2T's interaction design supports authors in fulfilling each of these responsibilities (2-5), with the exception of (1): building tutoring interfaces.

### 3.1 Preparing a Tutor Interface

Authoring-by tutoring begins from blank HTML interfaces that authors have prepared for their tutoring system. Among existing ITS authoring tools, there are several good options for purely graphical drag-and-drop-based interface builders. The CTAT HTML editor is one example [1], and in our user studies (section 6) we provided authors with interfaces prebuilt with this tool. Recent approaches also enable the auto-generation of interfaces from natural language descriptions [7]. AI2T can work with arbitrary HTML interfaces, so it is not particularly sensitive to the authors' interface authoring method, although in this work we limit the interfaces to only use buttons and text boxes.

### 3.2 Visualizing and Assisting Demonstration Explanations

After an interface is loaded into AI2T, authors begin building a tutor by filling in a start state for a single initial problem. Next, the author demonstrates a solution to this initial problem. AI2T induces primitive skills that can reproduce each of the author's demonstrated actions, and are later used by AI2T to solve new problems. To initially learn new skills AI2T self-explains each demonstrated action by searching for compositions of primitive functions (from a library of primitives) that can reproduce each demonstration. In this work, we limit this library to just a handful of arithmetic primitives that suffice for authoring the two domains in our user studies.

When self-explaining an author's action, AI2T may come up with just a few explanations or sometimes thousands. In either case, it is time-consuming to check and select among candidate explanations, so we enable the user to directly clarify the arguments and operations of the intended formula by two additional methods. The user can clarify the arguments they used to compute a demonstrated value by simply selecting interface elements on the screen. When the user writes in a demonstration they enter an *argument selection mode* indicated by replacing their mouse with a multi-colored cross-hair that toggles selected arguments when clicked (Fig. 4.a below). The selected arguments constrain the search for an explanation. Selecting arguments very often reduces the set of possible explanations to just one correct explanation. Users also have the option of stating the formula directly in natural language or in mathematical notation (Fig. 4.b below). These natural language explanations are mutually disambiguated with the demonstration using the method we describe in [58] to further narrow down the candidate explanations.

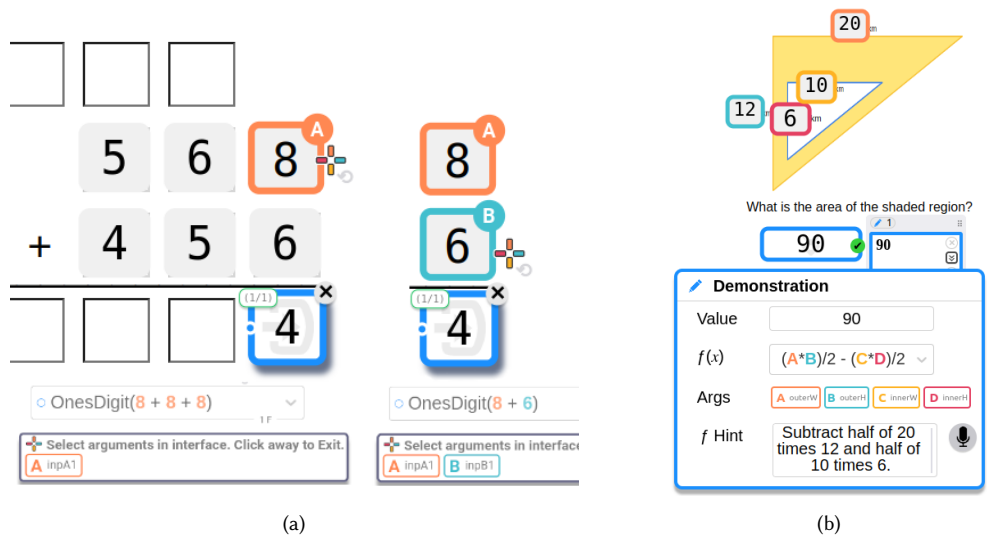

(a)                                                                              (b)

Fig. 4. (a) The user selects the interface elements that they used as arguments to compute the demonstrated 4. They first select the 8 and then the 6, and the agent's explanation for how the 4 was generated is immediately updated. (b) The user describes the formula that produced their demonstration 90 in natural language, and this is interpreted by the agent to guide the search for a formula that explains it.

Under the hood when AI2T produces a function composition it accounts for details such as extracting interface element attributes and converting between numbers and strings. For instance, to induce a skill that multiplies two numbers the agent must come up with a function composition like string(float(A.value) * float(B.value)). Where float() and string() convert between numbers and text strings, and A.value retrieves the string in the *value* slot of an interface element bound to variable A.

To simplify the explanation options displayed to users we drop all conversion operations and attribute de-references, reducing each formula to its simplest form (i.e. A * B), and then express them in terms of color-coded argument values instead of variables: 2 * 3. This color coding is mirrored in the interface by highlighting the borders of each argument interface element with corresponding colors (Fig. 3.a). This approach makes it clear what values are being used to explain the author's demonstrated action and resolves ambiguities between repeat values. If the agent comes up with multiple explanations for the same demonstration then they appear in a drop-down (Fig. 3.b). Mousing over each item in this drop-down highlights the arguments for one formula option in the interface, and clicking selects a formula as the correct one.

Fig. 5. The skill application window shows each of the actions proposed by the agent. Currently, action 3 of 4 is selected. The proposed actions convert the expression 5/6 + 2/3 to have a common denominator 18.

### 3.3 Supporting Completene Correctness Feedback for Each Problem State

After AI2T has induced skills from the author's demonstrations, it will try to apply those skills in new problem instances and propose actions that it believes are correct. While an author may train AI2T on 10-20 problems in a typical authoring session, they most likely will have only needed to solve two or three of those problems completely by hand. An author may spend about 5 minutes demonstrating solutions, and the remaining 15-25 minutes simply giving AI2T correctness feedback that helps it refine its skills so they are applied correctly in unseen problems.

Prior authoring-by-tutoring work [54] has reported usage patterns where authors validate just the first correct action suggested by the agent, but neglect to give feedback to other proposed actions in the same problem state. For getting an agent to solve problems with high accuracy (i.e. always produce one correct solution path) this is not a bad strategy. However, for authoring purposes we want the agent to be able to track all possible correct ways of solving problems. In this case, we aim to achieve 100% model-tracing completeness which means the agent should suggest all correct next actions and no incorrect actions for every reachable problem state. When users do not give feedback for all proposed actions, or neglect to demonstrate alternative correct next actions, the agent is deprived of important feedback toward achieving model-tracing completeness.

AI2T supports authors in recognizing when multiple actions are proposed—something that we struggled to design toward effectively in early versions of AI2T. The skill application window (Fig. 5) is the most important interface feature in this regard because it provides a visualization that lets authors sequentially flip between each of AI2T's proposed actions. Authors can flip through each action in the skill application window by selecting or hovering over each item. Clicking the ✗ or ✓ icons on the toggler (Fig. 5) in the skill application window assigns negative or positive feedback to the action.

Another way authors can give correctness feedback is by following the prompts to the right of the skill application window. The interface asks: "Is this action correct?". To which authors can respond by clicking the Yes or No buttons to give positive and negative feedback. After pressing Yes or No the next skill application is selected for feedback, and this repeats until all of the skill applications have been given feedback. After clicking Yes/No for the last skill application that needs feedback, one of the correct skill applications is selected and the prompt is displayed: "Demonstrate any other actions for this step. Press **Move On** to apply these actions". This encourages authors to consider whether AI2T has proposed every correct next action, and if not, demonstrate any missing actions. Authors can demonstrate actions at any time. There is no separate demonstration mode (e.g. like in SimStudent's interaction design) [39].

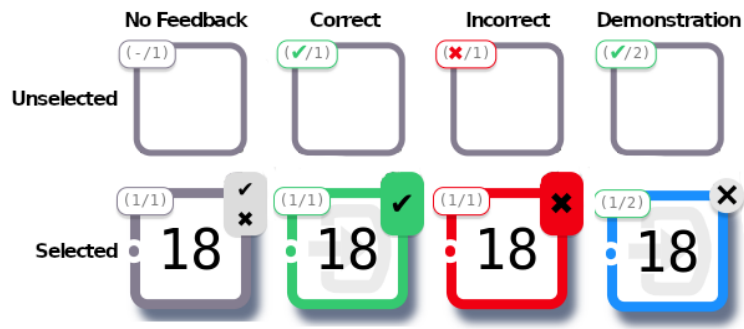

Fig. 6. Interface element overlays have an action count indicator when one or more proposed actions apply to that element. (Top) none of the actions are selected. (Bottom) one of those actions has been selected. The indicator and border are colored grey (left) when none of the skill applications have been given feedback, green (middle-left) when positive feedback has been given, red (middle-right) when negative feedback has been given, and blue when the selected skill application is a demonstration (right).

Each of AI2T's proposed actions also shows up as small indicators in the top-left of each interface element. The indicators show the number of proposed actions for each interface element (Fig. 6). Clicking on an interface element that has this indicator selects the first of its skill applications. These indicators are helpful for getting a quick sense of what interface elements the agent has proposed actions on, and which actions have been given feedback. The indicator's border is colored grey with a dash when no feedback has been given, green with a ✓ if any positive feedback has been given, and red with an ✗ if all of the provided feedback is negative.

When an action is selected this indicator shows the index number of the selected action. Also when an action is selected either a small toggler is shown next to its interface element to allow the user to quickly toggle positive and negative feedback, or if the action was produced by a demonstration then a small x button is shown to allow the user to remove that demonstration. This x button also shows up to the right of demonstrated actions in the skill application window.

### 3.4 Supporting Solution Path Navigation with Behavior Graph Generation

AI2T automatically generates behavior graph-like visualizations that help authors see the solution paths that they have trained AI2T on for each problem and navigate between different problem states. Since an AI2T agent learns hierarchical rule-like knowledge structures (not graphs), each generated behavior graph is simply a visualization of the agent's induced program applied to a particular problem—not a direct visualization of its internal knowledge structure.

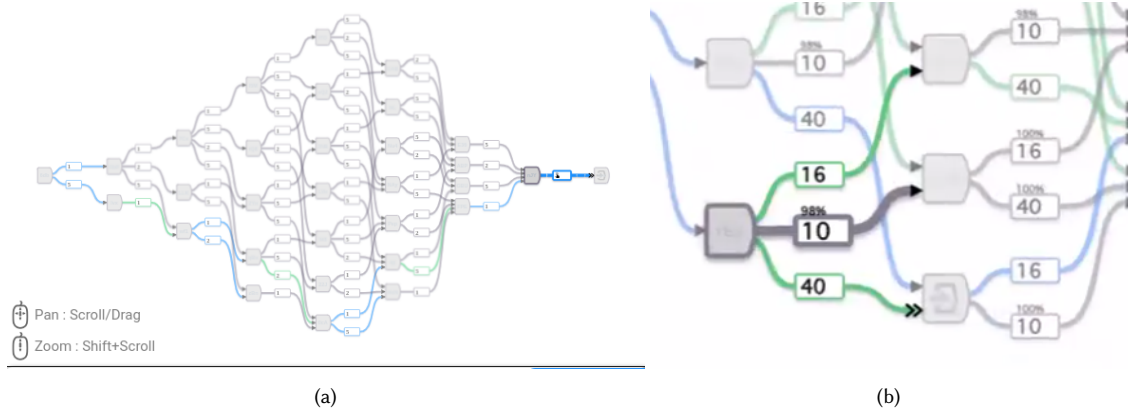

(a)                                                                                      (b)

Fig. 7. (a) A generated behavior graph and (b) a behavior graph zoomed into the current state. Of 3 proposed actions, the 1st and 3rd have been given positive feedback and the 2nd is selected. Demonstrations are blue, actions with positive feedback are green, and grey edges are proposed actions that still need feedback.

The behavior graph shows all of the diverging action sequences the agent believes are correct, and allows the author to navigate between these possibilities by clicking on particular nodes (for states) and edges (for actions).

To assist the author in tracking which paths have been tutored and which paths have not, each edge of the behavior graph is labeled with the value of its action and given a color indicating its feedback state. Blue edges indicate actions that the author demonstrated to the agent, grey edges indicate proposed actions that have not yet been given feedback, and green and red edges indicate actions that have been given positive and negative feedback respectively. We utilize a pallet that uses a particularly light green, dark red, and slightly blue-tinged grey, to reduce mix-ups common to various forms of color-blindness. We verified this pallet using an application that simulates protanopia, deuteranopia, and tritanopia and checked with some colorblind labmates.

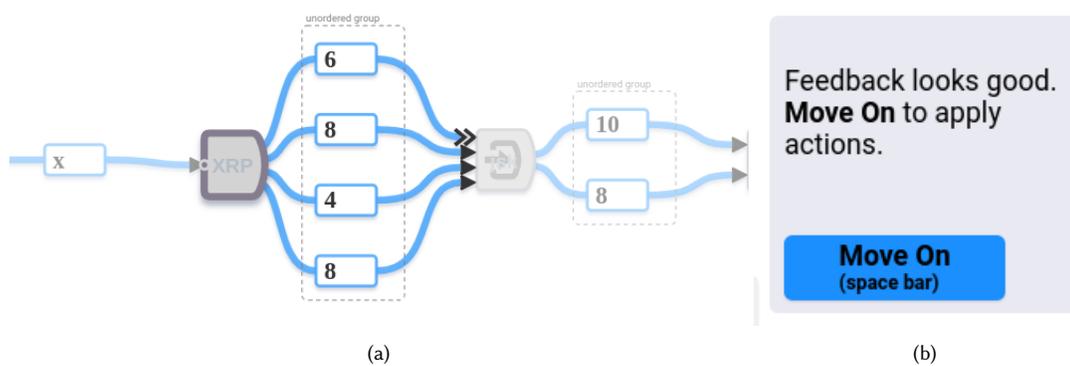

(a)                                                                                      (b)

Fig. 8. (a) A behavior graph with two visible unordered groups. (b) The Move On button moves the author to a new state by applying a single action or multiple actions in an unordered group.

Users can pan through AI2T's generated behavior graph by scrolling or dragging, and zoom in and out with shift+scroll. Entering a new state for any reason including clicking on a node, selecting an edge in a different state, or applying an action in the main interface, automatically animates the graph view so that it is centered on the new state and its downstream actions. This feature keeps the graph aligned with the current problem state. In piloting and both user studies users did not generally have much trouble with the mechanics of navigating through the behavior graph.

During user testing during the development of AI2T, we found that users struggled to understand behavior graphs when there were multiple solution paths, with the same actions but applied in different orders. To resolve this issue we introduced a feature where AI2T explicitly learns unordered groups as part of hierarchical task network induction (described later in section 4.1). These groups are displayed within the behavior graphs, considerably simplifying their structure so that one edge in the graph tends to correspond to just one unique action.

Unordered groups make the graphs easier to understand, but also save users time. Before AI2T could explicitly induce and display unordered groups, we found that users tended to apply feedback to all of the combinatorial solution paths produced by different orderings. They wanted to visit and grade the correctness of each edge. With the addition of unordered group induction, we made it so that by default multiple demonstrations applied simultaneously in the same problem state will create an unordered group. When the "Move On" button is applied all actions in an unordered group are applied together. This reduces the number of intermediate tutor states that authors need to visit and give feedback on.

### 3.5 Visualizing Action Certainty

Finally, within our revised skill application window, and on each edge of the behavior graph visualization, we added a continuous certainty score ranging between -100% and 100%. These scores indicate how sure the agent is that each proposed skill application is correct or incorrect. Negative values indicate that the agent is mostly certain that an action is incorrect and positive values indicate varying degrees of certainty that the proposed action is correct. As we describe in section 4.2, these values come from STAND's *instance certainty* measure which we will show is a fairly reliable indicator of prediction certainty and a good indicator of AI2T's actual learning progress—something we show is not true of many alternative methods of estimating prediction probability. Roughly speaking, if the agent proposes only actions with 100% certainty scores for a particular problem state, then this is a fairly strong indication to the author that the agent will exhibit 100% model-tracing complete behavior in similar situations. Mixtures of lower certainty scores are a fairly strong indication that the user should continue to train the agent on more problems.

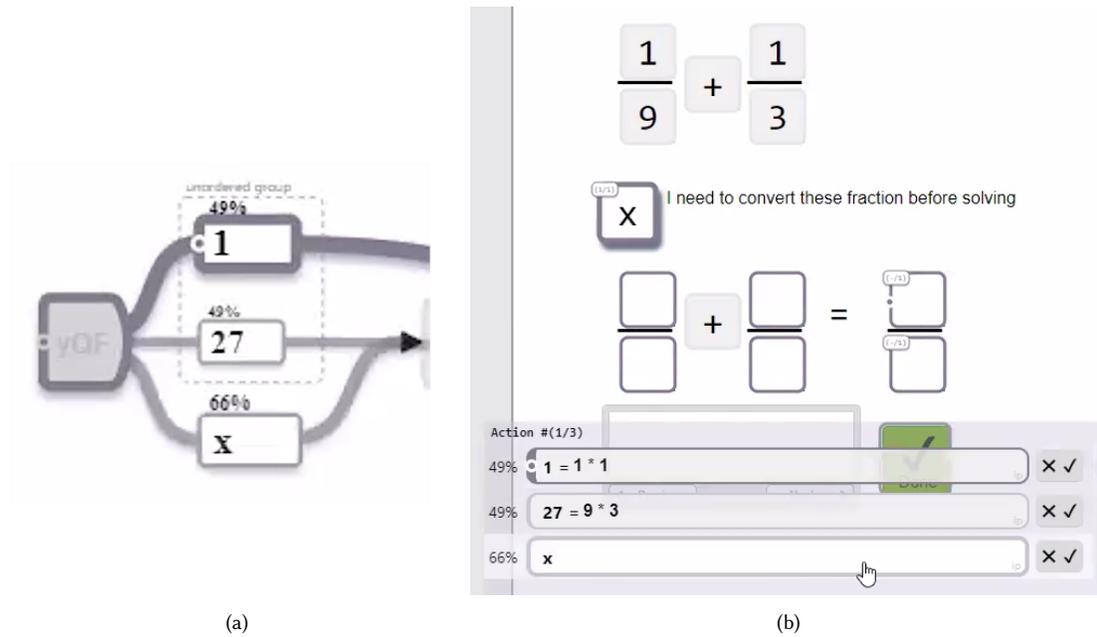

(a)                                                          (b)

Fig. 9. Three actions proposed for the initial state of 1/9+1/3. Certainty scores are shown over edges in the behavior graph (a) and in the skill application window (b). The first proposed action (49% certainty) would fill the numerator with 1=1*1. The first action is selected (indicated by a small white circle), but the third action is hovered over and previewed in the interface. This is the correct next action and has a higher certainty of 66%.

## 4   AI2T: Mechanisms for Self-Aware, Data-Efficient, and Robust Induction

Prior authoring-by-tutoring approaches have used simulated learners that simulate human-like induction [59] from demonstrations and supervised correctness feedback. These simulated learners largely share a similar breakdown of 3 core learning mechanisms that enable data-efficient induction that can be taught interactively. These mechanisms collectively induce several production-rule-like skills, that execute step-by-step solution strategies and flexibly track students' solutions in an ITS. Unlike hand-programmed production rules, skills are refined over the course of interactive training to reflect the author's instructed behaviors.

Each of the typical 3 core mechanisms in simulated learners used for authoring-by-tutoring induces different kinds of generalizations within each skill: 1) compositions of primitive functions that express *how* skills produce actions from other information in an interface, 2) patterns or concepts that capture *where* each skill might locate candidate inputs and outputs, and 3) preconditions that express *when* as in what contexts a potential candidate application of a skill is a correct application of the skill. The mechanisms that learn these different types of generalizations are typically referred to as *how-*, *where-*, and *when-learning* mechanisms respectively [39, 54].

AI2T's learning-mechanism implementations are similar to prior implementations of AL [54] with the exception of two important innovations. AI2T uses an algorithm called STAND for *when-learning* (i.e., precondition induction), and introduces a fourth learning mechanism which we call *process-learning* that organizes induced skills into hierarchical task networks (HTNs). Unlike methods that have users describe HTN structures directly in a top-down manner [23, 30, 34] AI2T learns HTNs directly from authors' demonstrated action sequences (i.e. via bottom-up induction).

Process-learning requires no new interactions from authors. Authors still only need to solve problems and grade AI2T as it solves problems. The author does not need to plan or verify any part of processs-learning's induced HTNs. In our user studies (section 6) we do not even display AI2T's induced HTNs to participants.

### 4.1 Process-Learning: Hierarchical Task Network Induction from Action Sequences

AI2T's induced HTNs recursively break tasks into subtasks that terminate in primitive action-producing skills. Methods in the HTN are higher-order skills that carry out tasks as ordered or unordered sequences of sub-tasks and primitive skills. Figure 10 shows an example of an induced HTN for an ITS (like Figure 10) that supports students in practicing and deciding between different fraction arithmetic procedures like adding, multiplying, and conversion. For instance, adding fractions with unlike denominators first requires converting them to have the same denominator (method 3), but multiplying fractions (method 2) or adding fractions (method 4) with equal denominators does not require conversion. In the HTN the choice between these three options is represented by a disjunction (an OR) where three methods share the same parent task "Combine Fraction Expression". Each method's induced preconditions (from when-learning) gate their consideration as acceptable solution strategies. In this example, the correct preconditions would make these methods mutually exclusive. However, in problems with multiple valid approaches to achieving the same subtask, multiple methods may be applicable simultaneously. For instance, Figure 10 captures just one method (#5) for the "Convert Fractions" subtask, but an author could certainly add another to the HTN by demonstrating it at the appropriate problem step.

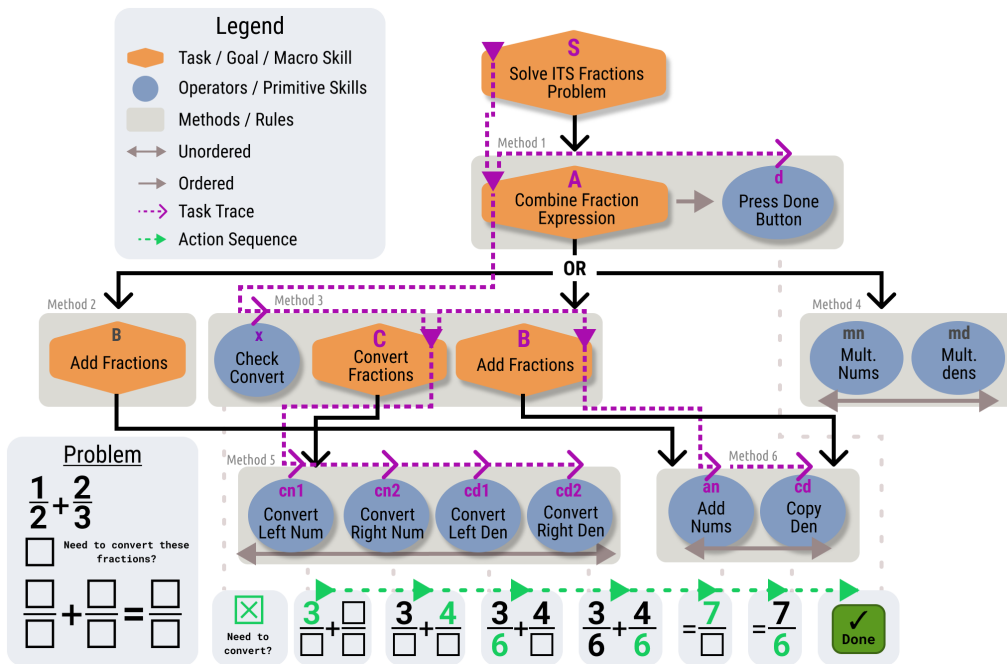

Fig. 10. A possible HTN for a fraction arithmetic ITS. A task trace (dotted-purple) and action trace (dashed-green) are shown of the solution to an addition problem of fractions with different denominators.

Learning correct ordering constraints with the typical 3-mechanism approach (like SimStudent and AL) requires the author to give agents negative feedback when they suggest applying individual skills out-of-order; often across dozens of problems. For instance, prior authoring-by-tutoring agents built with AL [54] have struggled to strictly sum multi-digit numbers from right to left until authors first corrected several out-of-order mistakes. Inducing skill preconditions with the typical 3-mechanism simulated learner approach often requires collecting diverse sets of negative examples.

Adding process-learning (for a 4-mechanism approach) considerably simplifies the learning of ordering constraints. Process-learning situates primitive skills into a higher-order procedural process, and so solution ordering begins as strictly sequential when AI2T has only seen one solution sequence, and ordering constraints are relaxed as examples of solution flexibility arise throughout training. Solutions to new problems and demonstrations of unordered actions within the same problem can generalize the HTN and introduce new disjunctions or unordered groups.

An important feature of AI2T's process-learning mechanism is that its HTN induction is agnostic to lesson order. Prior simulated learner approaches that induce HTNs, like VanLehn's Sierra system [51, 52], have required curated lesson sequences or suffer complete inductive failure. By contrast, AI2T will induce the same HTN structure regardless of the order in which it experiences new problems, meaning authors do not need to explicitly design good lesson plans for AI2T.

AI2T's HTN representation language also allows for the induction of HTNs with optional or conditional symbols in methods. For instance, carrying a 1 in a multi-column addition problem is a conditional step—it only occurs for partial sums greater than 10. An example of an optional step may be writing the numbers that are added or subtracted from each side of an algebra equation before calculating the next line—an author may not want to penalize students who explicitly skip this step, but still track student solutions and provide adaptive supports in cases where that step is not skipped. Similarly, recent work has promoted HTN-based model-tracing as a favorable ITS design paradigm [50]. HTNs naturally capture knowledge at multiple levels of granularity, and thus are a good knowledge representation for ITSs that support adaptive scaffolding where incremental step-by-step tutoring support is faded as students gain greater proficiency [25, 41, 49]. We do not evaluate this particular feature in this work, but AI2T's HTN induction approach is certainly conducive to supporting it.

### 4.2 STAND: Self-Aware Precondition Induction

In prior work, when-learning has been identified as a major limiting factor for efficient authoring. Typically how- and where-learning converge to their final generalizations from just one or two examples. When-learning requires several additional correct and incorrect examples to induce correct generalizations. This is true even with the inclusion of process-learning, although process-learning provides structure that can considerably simplify the preconditions that skills need to induce to operate properly.

STAND provides more data-efficient precondition induction than prior when-learning approaches [57]. Prior approaches include fitting decision trees [37, 54] or applying inductive logic programming methods [39] to learn each skill's preconditions. Instead of learning a single generalization to predict if a candidate application of a skill is correct STAND learns a space of classifiers consistent with the authors' positive and negative training examples. STAND accounts for a complete set of *good* candidate generalizations instead of selecting a single generalization by breaking ties randomly. In this way, STAND explicitly models the inherent ambiguity of trying to learn generalizations that perform well on unseen examples—something that is especially challenging when there is limited training data.

STAND induces spaces of generalizations that are structurally similar to version spaces. However, STAND suffers from none of the drawbacks of the typical candidate elimination approaches for version-space learning [42]. STAND does not suffer from version-space collapse where induction fails completely in response to noise. STAND can also learn full disjunctive normal logical statements instead of just strictly conjunctive ones. When used within AI2T these generalizations consist of relational literals (with variables), because AI2T prepares features for when-learning so that they are restated relative to the selection variable (the interface element being acted on) and argument variables (elements being drawn from), which are defined by each skill's matching pattern (induced by *where-learning*). Finally, STAND is very computationally efficient, meaning it does not add additional lag time between authors' interactions. STAND is only marginally slower than decision tree learning, yet it essentially learns every *good* decision tree that fits a dataset. This speed is achieved by a clever form of data structure compression.

In addition, STAND can produce a measure (from -100% to 100%) called *instance certainty* [57] that estimates how certain it is of predicting the correctness label of an unseen example. STAND's instance certainty on single unseen examples tends to increase when its holdout set performance increases, meaning it is a good heuristic for actual learning gains. Conventional methods of estimating prediction probability over other data-efficient induction methods, including ensemble methods like random forests and XGBoost, simply lack this property (we demonstrate this in the next section). *Instance certainty* is more effective than these alternatives because it captures how unambiguous the label prediction of an example is given all of the generalizations that capture the example. Low instance certainty indicates high disagreement between the predictions of alternative generalizations in STAND's version space. STAND's benefit over conventional ensemble methods is that it captures the predictions of a space of all *good* generalization candidates instead of just several randomly selected ones. When STAND learns the correctness label of a low instance certainty example, it will tend to increase its holdout performance considerably, because the example will cut out many bad generalizations still captured by its approximate version-space.

## 5   Simulation Experiments

In addition to testing AI2T with users, we wanted to ensure that STAND and process-learning produce improvements in learning efficiency over prior approaches. In these experiments, we first evaluate STAND using it for when-learning in a typical 3-mechanism simulated learner configuration (with how-, where-, and when-learning, but not process-learning). We compare STAND to various alternative when-learning approaches using an automated training system that mimics the demonstrations and feedback that an ideal user would provide while authoring. We apply this special *authoring* training approach in the two domains that we had participants author in our user study (section 5): multicolumn addition and fraction arithmetic.

In this setup, each agent receives ideal on-demand demonstrations and correctness feedback. At each state, all proposed actions are given correctness feedback. If an action is missing then it is demonstrated to the agent with annotations that make the underlying reason for the action unambiguous. Each demo is annotated with the formula for producing the action's value, and the arguments used. This replicates the behavior of an ideal user who always selects the correct formula to explain each demo (among the several suggested possibilities). These annotations enable how- and where-learning to produce error-less generalizations almost immediately, meaning almost all errors can be attributed to when-learning. No annotations are provided to assist when-learning besides the correctness labels of each action. Just like an ideal author, the training system trains the agent on all alternative solution paths for each problem.

We compare several classifiers with STAND:

(1) **Decision Tree**: A decision tree using gini impurity [6] as the impurity criterion. We use STAND's implementation, expanding just one random split at each decision point.

(2) **Random Forest**: Scikit-learn implementation of random forest ensemble [5] of 100 decision trees. Random forests use bagging [4] to independently train several decision trees on subsets of the data.

(3) **XG Boost**: An ensemble method that trains multiple decision trees one at a time. This method uses gradient-based sampling to re-weight the samples for subsequent trees [10].

These tree-based methods are chosen because they excel at learning from small datasets of structured data. In all models no limits are set on tree depth or leaf size since for these condition-learning tasks the available features from the tutoring interface are sufficient for separating correct and incorrect candidate skill applications perfectly. Since condition learning is noiseless the trees will already tend to not become more complex than the ideal solution, and limiting their depth could only prevent the ideal solution from being discovered.

The two ensemble methods are included for comparison with STAND's comprehensive version-space-based approach, and to compare the utility of their prediction probabilities with STAND's instance certainty estimates. Each model is re-trained on 40 repetitions on a sequence of 100 randomly generated problems.

### 5.1 Productive Monotonicity: Certainty Score Change vs Holdout Set Performance Change

Productive monotonicity is defined as the proportion of changes in certainty estimates for actions in a holdout set that move toward 100% when the action is correct and -100% when the action is incorrect. High productive monotonicity reflects the degree to which changes in certainty estimates mirror actual learning gains (increases in holdout set performance).

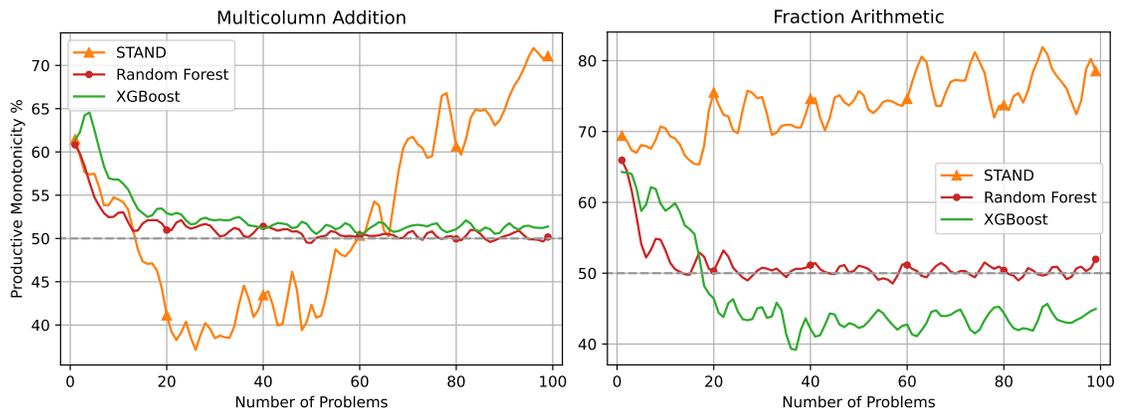

Fig. 11. Productive Monotonicity By Problem

Our results show that STAND's instance certainty measure has considerably higher overall productive monotonicity than the two ensemble methods' prediction probabilities. Thus, increases in instance certainty reflect actual increases in holdout set performance, meaning instance certainty is a relatively good heuristic for determining when STAND has learned. By comparison, the random forest and XG Boost's prediction probabilities align with actual changes in holdout performance only about 50% of the time—they are not much better than chance.

In multi-column addition, STAND's productive monotonicity is < 50% for the first 60 training problems and > 50% thereafter. This pattern may occur in this domain because in the early stages of training STAND's approximate version space is still growing from entertaining new disjunctions, growing the space of generalizations that STAND considers. Fractions may not show a similar pattern because purely conjunctive preconditions tend to suffice in this domain, and so STAND's effective version space tends to shrink monotonically throughout training.

## 5.2 Precision at High Certainties

If a when-learning classifier predicts that an action is correct with a high certainty of 90%-100% then there should be a very low probability that the action is actually incorrect.

Table 1.  Total Precision at High Certainties

|  | MC Addition | | Fractions | |
| --- | --- | --- | --- | --- |
|  | ≥ 90% | = 100% | ≥ 90% | = 100% |
| STAND | 93.19% | 99.81% | 95.70% | 100.00% |
| Random Forest | 97.15% | 94.79% | 95.19% | 93.72% |
| XGBoost | 98.35% | 100% | 99.39% | 100.0% |

Our simulations show that XGBoost has the highest precision at high certainties. For predictions of 100% STAND is nearly as precise as XGBoost in multicolumn addition and equally 100% precise in fractions. For predictions of ≥ 90% STAND's precision is closer to 90%, which is arguably a desirable property—as it indicates some alignment of instance certainty with actual ground-truth precision. These results validate that certainty scores of 100% are typically only given to actions that are truly correct, and that certainty scores between 90% and 100% tend to apply to actions with a small probability of error.

## 5.3 Per-Problem Completeness

Finally, we verify that STAND and process-learning produce more data-efficient learning and higher rates of 100% model-tracing completeness on holdout data. We report each model's model-tracing completeness on a holdout set of 100 problems, evaluated at the end of each training problem.

In both domains, STAND's average completeness is higher than the competing models throughout the training sequence. This implies that STAND has better data efficiency and asymptotic performance since it can achieve greater levels of completeness with fewer training problems. In 19 of 40 MC addition repetitions STAND achieved 100% completeness after training on a sequence of 100 problems compared to 10 of 40 repetitions for decision trees. In fractions, 38 of 40 repetitions achieved 100% completeness with STAND and decision trees. The relative performance of the decision tree, random forest, and XG Boost varies between domains. Notably the random forest was the worst in multicolumn addition, likely because its bagging approach of sampling subsets of the data had the effect of dropping important edge cases, which are particularly important in this domain.

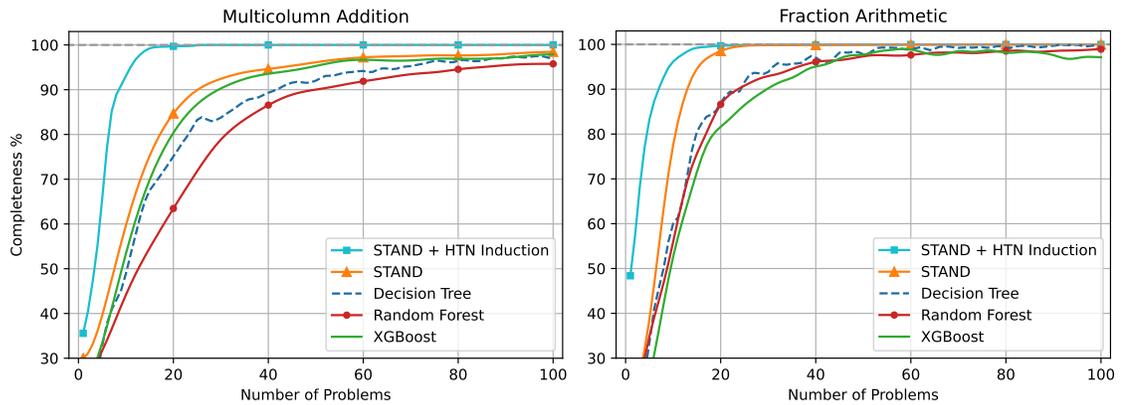

Fig. 12. Average holdout completeness by problem.

Table 2. Average Holdout Completeness at Problem N, and Number of 100% Complete Repetitions at problem 100.

|  | Multicolumn Addition | | | | Fractions | | | |
|---|---|---|---|---|---|---|---|---|
|  | N=20 | N=50 | N=100 | 100% Reps | N=20 | N=50 | N=100 | 100% Reps |
| STAND + HTN Induction | **99.72%** | **100.0%** | **100.0%** | **40/40** | **99.69%** | **99.94%** | 99.99% | 38/40 |
| STAND | 85.45% | 96.10% | 98.62% | 19/40 | 98.72% | 99.91% | 99.99% | 38/40 |
| Decision Tree | 75.75% | 91.86% | 96.97% | 10/40 | 88.15% | 97.24% | 99.88% | 38/40 |
| Random Forest | 64.16% | 90.02% | 95.53% | 0/40 | 88.13% | 97.44% | 98.97% | 11/40 |
| XG Boost | 81.20% | 95.40% | 98.01% | 3/40 | 81.12% | 96.20% | 97.34% | 27/40 |

Introducing process-learning so that AI2T induces HTNs instead of independant primitive skills improves AI2T's learning even further. In this configuration, AI2T's absolute holdout set performance is strictly higher than the competing models throughout training, and in every single training repetition 100% model-tracing completeness is achieved in multicolumn addition. These results validate the absolute performance benefits of STAND and show empirically how process-learning promotes model-tracing completeness beyond what can be achieved with the typical 3-mechanism approach used by prior authoring-by-tutoring systems.

## 6 User Studies

We evaluated AI2T in two studies each with 10 users in which participants authored tutoring systems for multicolumn addition and fraction arithmetic. In each domain participants tutored an AI2T agent on several problems until they were convinced that the agent could produce correct and complete behavior for any new problem instance. When participants self-reported that they believed the agent had achieved a state of absolute completeness, we scored their agents' model-tracing performance on a large holdout set of 100 problems. After being scored, participants moved on to the next domain.

The central aim of these studies was to evaluate what configurations of the agent and interface design best support authors in teaching AI2T to induce correct and complete programs. Beyond qualitative observations of usability, a core element of this evaluation is to determine whether our interaction design supports authors in building model-tracing complete tutoring systems.

Very little prior work explores interactions for supporting users in self-assessing the learning progress of interactively teachable AI agents. Explainable AI methods, employed for this purpose [16] typically apply to conventional probabilistic predictors trained over large datasets. In these cases, interactive inputs from users like feature suggestions offer marginal benefits to models that will likely never achieve perfect performance in their target task. AI2T is among rarer cases of AI agents that succeed at bottom-up induction of trustable programs from interactive instruction over arbitrary HTML interfaces (not specialized environments) [14, 48]. Our two studies evaluate not just how well AI2T's design supports users in engaging in effective training, but also how well its certainty score feature supports them in determining when they have trained 100% model-tracing complete programs.

## 6.1 Methods

*6.1.1 Participants.* All users participated in these studies remotely via Zoom, and screen shared as they worked in AI2T's web interface. Participants filled out online forms indicating their consent to be recorded. For their participation in these IRB-approved studies, users were compensated with a $30 Amazon gift card. We limited all sessions to a maximum of 90 minutes. Participants for piloting prior to study 1 included labmates and colleagues who had volunteered their time. The 10 paid participants for study 1 were all graduate students recruited from Carnegie Mellon University (CMU) and 8 of these 10 participants were part of graduate programs that specialized in educational technology. 4 more CMU graduate students were recruited and compensated as part of piloting prior to study 2. The participants for study 2 included 5 graduate students from CMU, 3 of which specialized in educational technology, 2 human-computer interaction graduate students, 4 biology graduate students from Arizona State University (ASU), and 1 professional specializing in the authoring of instructional technology for a major ITS project unaffiliated with CMU or ASU. Participants' self-reported genders were roughly equally male and female in both studies.

The typical end-users for an authoring tool like AI2T include instructional designers, learning engineers, teachers, and researchers. Our participant population of mostly graduate students is fairly well aligned with the educational backgrounds of these populations which typically have some graduate education. Many of our participants explicitly study the design of educational technology (although not necessarily the programming elements of it). All participants have bachelor's degrees, and 6 of the 10 participants in study 2 were non-programmers. Several of the study 1 participants indicated that they were non-programmers, but we did not collect this data systematically in study 1.

For study 2 we asked participants to score their programming experience on a Likert scale from 1 to 5. We asked: "Would you describe yourself as a proficient programmer (i.e. you have the ability to write scripts/software)" where 1 is labeled as "No. I have little or no programming experience.", and 5 is labeled as "I have extensive programming experience. I believe I could program professionally." For the purposes of our analyses, we consider a score of 1 or 2 to be a non-programmer.

*6.1.2 Instruction.* In both studies, participants authored multicolumn addition before fraction arithmetic. None of the participants had used AI2T before, so in both studies, we can think of this first domain as a warm-up attempt to practice using the tool. We first gave participants a short tutorial on how to use the tool by showing them how to demonstrate each step of the problem 777+777, and showed them how to give feedback to the agent on the subsequent problem 222+222. Prior to having participants begin authoring each domain, we described the behavior that we expected the final tutoring system to have, and we asked that they engage in a think-aloud: "Say whatever you are thinking as you work with the tool". Participants always began authoring with a blank agent with no prior training. Participants were provided a blank interface for each domain, they did not make their own.

Fig. 13. Study 1 Interface. An intermediate state of problem 597+346. Two of three actions (outgoing graph edges in top-left) have been given negative feedback (marked red). A third action (grey) is selected, it adds 7+5 and takes the ones digit resulting in the value 2. Pressing the Yes button will mark this selected action as correct.

While users worked with AI2T we made ourselves available to answer questions and made participants aware if they began to teach the agent a procedure that differed from the target ITS behavior. We refused to give participants any advice concerning when they should stop training the agent. We limited ourselves to suggesting that they train the agent on a variety of problems and suggested that they should at least keep training AI2T until it seemed like it had stopped getting problem steps wrong. We never explicitly pointed out the availability of certainty scores in the interface. We wanted to assess whether users noticed, understood, and utilized these indicators on their own.

At the end of each session, we asked users to give their overall feedback about the system. We simply asked: "What worked well and what didn't work well as you were using the tool?". In study 2 we also explicitly asked participants if they noticed the certainty score indicators and whether they considered them when deciding whether or not to stop training the agent.

## 6.2 Study 1: 3-Mechanism Agent

Study 1 was conducted to evaluate the efficacy of an in-development version of AI2T's prior to the implementation of process-learning. Study 1 was not particularly successful, but we share it here because the contrast between studies 1 and 2 highlights the importance of certain features that help support intuitive and effective authoring-by-tutoring.

Three things are different in the study 1 version of AI2T 1) the agent has no process-learning mechanism, 2) the behavior graph does not display unordered groups so the graphs branch combinatorially, and 3) there is no skill application window so authors must use the edges of the generated behavior graphs or the indicators in the tutor interface to locate and switch between different proposed actions. In this version, certainty scores only appear over edges in the behavior graph, since there is no skill application window.

As we have shown in simulation, the typical 3-mechanism approach (without process-learning) struggles to induce multi-column addition. To assess the study 1 interface's support for convergence to 100% model-tracing completeness we chose to have participants (in study 1 only) author a simpler *zero-carry* variant of multicolumn addition where a 0 is placed in each carry slot if a 1 would not normally be carried. This variant is easier for 3-mechanism agents to learn because most skills require simpler preconditions. When 0 or 1 is always written explicitly in the carry space, the agent can use this as a display-based indication for moving on to the next column. In the normal procedure, the agent must induce complex cross-skill preconditions that check that the carry step will be skipped in the previous column.

Table 3. Results for Study 1

|  | MC Addition (zero-carry) | | Fraction Arithmetic | |
|---|---|---|---|---|
| User# | Completeness | Minutes | Completeness | Minutes |
| 1 | 90% | 55 | - | |
| 2 | 99.80% | 33 | 64.69% | 32 |
| 3 | 71.43% | 31 | 100% | 24 |
| 4 | 100% | 33 | 92.76% | 34 |
| 5 | 91.45% | 24 | 100% | 33 |
| 6 | 90.60% | 24 | 38.05% | 32 |
| 7 | 100% | 22 | 85.75% | 35 |
| 8 | 100% | 23 | 95.29% | 23 |
| 9 | 99.15% | 28 | 76.43% | 29 |
| 10 | 57.07% | 50 | - | - |
| **Mean** | 90% | 30.33 | 82% | 30.25 |
| **Median** | 95% | 28 | 89% | 32 |

6.2.1  *Quantitative Results.* The quantitative results for study 1 are outlined in Table 3. For the *zero-carry* version of multicolumn addition, 3 of 10 participants taught agents that achieved a 100% model-tracing completeness score on the holdout set of 100 random problems. Two of our participants took more than half of the allotted 90 minutes for the first domain, and we did not have them complete the second. In fraction arithmetic just 2 of the 8 participants achieved 100%. In both domains, the total authoring time—the time between beginning authoring and self-reporting that they believed the agent had achieved correct and complete behavior—was about 30 minutes.

6.2.2  *Qualitative Results.* Automatic behavior graph generation is one major improvement in AI2T over prior authoring-by-tutoring interaction designs [54]. Users generally had little trouble panning and selecting states and actions in the behavior graph. However, some users had difficulty connecting patterns in the behavior graph with our instructions that certain subsets of actions should be permitted in any order. Without unordered group induction, alternative action orders are displayed as diverging paths. A key element of users' difficulty was that actions essentially needed to be demonstrated or given feedback multiple times since they could appear as distinct edges along different paths.

For the fraction arithmetic domain, some users found it tedious to give feedback to the agent in 24 unique problem states generated from each permutation of the 4 steps associated with converting two fractions. In reality, users may have needed to only give feedback on a small subset of these states per problem to achieve 100% completeness. However, since the behavior graph made it very easy for users to see what states they had and had not given feedback on, and since we had set them upon the objective of 100% completeness, most participants tended to grade the agent along all generated paths.

Behavior graphs also showed certainty scores above each edge in the behavior graph. However, this presentation did not appear to be effective as none of the users mentioned the certainty scores in their think-alouds or follow-up interviews, and those that we asked explicitly said they did not notice them.

Many of the participants in study 1 had used other ITS authoring tools before, including CTAT's example-tracing authoring tool. In follow-up interviews, several participants commented that they found AI2T easier to use than CTAT example-tracing because after demonstrating solutions to a single problem, the agent would suggest step-by-step solutions automatically for the remaining problems. For instance, one participant remarked "This is a lot nicer than CTAT… I like that it mostly does the problems for you." These users commented that checking the agent's step-by-step solutions was much easier than demonstrating several problems themselves, and noted that the fact that the agent induced a program from their teaching saved them from needing to mass produce step-by-step problem solutions in a spreadsheet.

### 6.3   Study 2: AI2T with Process-Learning

Fig. 14.  Normal AI2T interface (study 2) in an intermediate state of 189+542. There are 4 proposed actions, shown as skill application window items (middle-bottom) and graph edges (top-left). In the graph, they are divided into two unordered groups (dashed boxes). Two actions have been given negative feedback (✗), and one has been given positive feedback (✓). The mouse hovers over an action without feedback that was proposed with 88% certainty. When the ✓ is clicked on the toggle button it will be given positive feedback. When "Move On" is clicked it will move to the state after both the correct actions in the top unordered group are applied.

The configurations of both the backend agent and frontend interface differ between study 1 and study 2. Taken together they form a loose pseudo-experiment, in which study 1 establishes a baseline with several issues and study 2 implements several fixes to remedy those issues. The version of AI2T deployed in study 2 is as we have described earlier in section

3. By comparison, to study 1 the agent has a process-learning mechanism, shows unordered groups in graphs, and presents proposed actions in a skill application window.

Since participants were not randomly assigned to conditions and since there are multiple differences between the two studies we cannot precisely quantify the benefits of each improvement. However, as we will show in the following subsections participants in study 2 were far more successful at producing 100% model-tracing complete tutoring systems. Our simulation results would lead us to believe that this difference is largely due to the more rapid and robust learning that can be achieved with the inclusion of a process-learning mechanism. However, our qualitative observations and interviews with participants led us to believe our interaction design improvements played a large role as well.

*6.3.1 Quantitative Results.* Table 4 outlines the results for study 2. 6 of our 10 participants reported their programming experience as a 2 out of 5. These 6 non-programmer participants included our 4 biology graduate students, and 2 graduate students specializing in design.

8 of 10 participants succeeded at training agents that achieved 100% holdout completeness for multicolumn addition (the normal *non-zero-carry* version). Prior work in this domain had reported lower median model-tracing completeness rates of 92%, with no instances of 100% [54]. Our study 2 results also showed users completing training in about half the time compared to prior work: a median of 22 minutes instead of 41 minutes. For multicolumn addition, all users first solved the same 7 problems and selected their own problems thereafter. In most cases, two or three additional problems after the initial 7 were sufficient to achieve 100%. The two participants who did not reach 100% made mistakes during training that they did not succeed in tracking down and fixing.

Table 4. Results for Study 2

| User# | Prog. Exp. | MC Addition (normal) | | | Fraction Arithmetic | | | Notice | Use |
|---|---|---|---|---|---|---|---|---|---|
| | | Completeness | Minutes | N prob. | Completeness | Minutes | N prob. | | |
| 1 | 2 | 100% | 22 | 13 | 100% | 21 | 20 | | |
| 2 | 2 | 100% | 20 | 9 | 100% | 32 | 18 | ✓ | ✓ |
| 3 | 2 | 100% | 30 | 11 | 96.31% | 30 | 16 | | |
| 4 | 5 | 100% | 14 | 11 | 88.52% | 17 | 13 | ✓ | ✓ |
| 5 | 2 | 90.96% | 30 | 14 | 40.18% | 19 | 9 | | |
| 6 | 5 | 100% | 22 | 11 | 98.87% | 16 | 10 | | |
| 7 | 5 | 100% | 28 | 11 | 100% | 27 | 14 | ✓ | ✓ |
| 8 | 3 | 89.16% | 36 | 11 | 38.73% | 18 | 14 | | |
| 9 | 2 | 100% | 22 | 10 | 100% | 25 | 18 | ✓ | ✓ |
| 10 | 2 | 100% | 18 | 9 | 100% | 23 | 21 | ✓ | |
| **Mean** | 3 | 98% | 24.2 | 11 | 86% | 22.8 | 15.3 | 5/10 | 4/10 |
| **Median** | 2 | 100% | 22 | 11 | 99% | 22 | 15 | | |

In fraction arithmetic, participants self-selected all of their own problems, and 5 out of 10 participants succeeded at training agents that achieved 100% model-tracing completeness. The two lowest-performing participants made mistakes during training that prevented them from achieving more than 50% completeness. The median completeness in this domain was 99%. Participants trained the agent on 9 to 21 problems in this domain in 16 to 32 minutes with a median of 22 minutes.

In our follow-up interviews we asked participants if they noticed the certainty score indicators, and whether they considered them when deciding when to stop training the agent. 5 of 10 participants said that they did notice the certainty scores, and 4 indicated that they considered them when deciding whether or not to stop training the agent. Specifically, these participants indicated that they took the presence of a low certainty action as an indication that the agent needed additional training on similar problems. 3 of these 4 participants achieved 100% model-tracing completeness in both domains.

*6.3.2 Qualitative Results.* In our follow-up interviews several participants remarked on how quickly they learned to use our tool, and how they could succeed at using AI2T to author two tutoring systems in less than an hour. As in study 1, several users remarked on how quickly the agent was able to learn from their instruction and how the authoring process became much easier once they entered the stage of mostly checking the agent's behavior on new problems. For instance, one of our biology graduate student participants remarked: "This is wild, I could teach it all that math in like 20 minutes [per topic]."

While some study 1 participants had commented on issues with the smoothness of the "interaction loop", very few study 2 participants had constructive negative feedback. Our observations of users led us to believe that the inclusion of the skill application window in the study 2 design was helpful in this regard. Some study 1 participants jumped between problem states without fully giving feedback to all of the agent's proposed actions, whereas study 2 participants very consistently fell into a pattern of looking through actions in the skill application window and giving them all feedback before moving on. The inclusion of unordered groups meant that there were far fewer states in the study 2 version for users to navigate through. For instance, much of users' time in study 1 was spent going through many diverging states in large behavior graphs, especially for the fraction arithmetic domain, but unordered groups spared study 2 users from the tedium of grading combinatorial paths. While some users in study 1 expressed that they had become disoriented navigating between problem states, study 2 participants did not indicate any similar issues.

Our follow-up interviews provided strong evidence that the users who noticed the certainty scores used them successfully to gauge when they should stop training the agent. For instance, participant 2 in study 2 said, "I definitely would have stopped teaching it earlier if I hadn't seen the low confidence on some problems that I thought it already knew how to do."

## 6.4 Discussion

Overall our study 2 results show that our redesign produced a considerable improvement over study 1. Half of the study 2 participants succeeded at training agents with 100% complete tutoring system behavior on both domains, usually in under half an hour. Our interviews with users also confirmed that displaying STAND's instance certainty measure was useful for assessing the AI2T agent's learning progress toward 100% completeness. Several participants in study 2 indicated that this indicator influenced their decision of when to stop training the agent on new problems. This is a strong preliminary indication that the certainty score indicators had the intended effect. A future randomized experiment would be able to lend stronger statistical evidence for the connection between the availability of this indicator and high authoring completeness. However, the productive monotonicity measure we report in our simulation experiments already establishes that this measure accurately reflects agent learning, so it is reasonable to conclude that if users were explicitly trained to interpret it, they could use it successfully as a heuristic for estimating holdout completeness.

In study 2, when users did not achieve 100% model-tracing completeness they either made clear mistakes during authoring (e.g. users 5 and 8) or trained AI2T on too few problems. Thus, improving AI2T's robustness may largely come down to better support for training users and helping them catch mistakes. Participants 3, 4, and 6 likely fell short of 100% because they trained AI2T on too few problems in fraction arithmetic (we did not identify any uncaught mistakes upon reviewing their screen recordings). Participant 4 was the only user among these three who claimed to use certainty scores, and the only participant familiar with the trajectory of the AI2T project—specifically that AI2T had become more data-efficient in the months prior to study 2—and thus they may have had a skewed belief of how little training was required. When participants strictly interpreted a less than < 100% certainty score as an indication of incompleteness they trained AI2T on a sufficient number of problems. Consequently, a little bit of training to check for mistakes, and correctly interpret certainty scores may go a long way toward helping authors use AI2T effectively.

## 6.5 Who can use AI2T?

The major aim of this work was to prototype a method whereby the authoring of complex ITSs is made simple, fast, and accessible to non-programmers. Instructional designers, learning engineers, teachers, and researchers are all professionals who may or may not have programming expertise, but all certainly benefit from being able to author complex ITSs quickly. Many of our non-programmer participants succeeded at authoring complete model-tracing behavior in less than half an hour, and thus there is every reason to believe that the same would be true for in-service professionals who regularly think about tutoring system design or grade student work.

# 7 Future Work

## 7.1 Design Support for Open-ended Authoring

Our focus of this work was AI2T's usability, but more open-ended authoring evaluations may shed light on the unique needs and design perspectives of in-service professionals. ITS authoring typically requires authors to deliberately design interfaces around adaptive step-by-step support, and this design process typically involves a design loop: a cyclic process of classroom testing and revision [2]. A common mistake in this process is designing instruction around final problem solutions or problem-solving with too large of steps that do not sufficiently break down strategies into their most fine-grained elements. AI2T may play a beneficial role in supporting more adaptive designs earlier in development. Much like students with low-prior knowledge who benefit from bite-sized instruction that reduces cognitive load [25, 41, 49], AI2T tends to learn more effectively from granular step-by-step instruction. Prior work has demonstrated that this feature of inductive simulated learners can be useful for automated student model discovery (for knowledge tracers) [32], and as a tool for cognitive task analysis [38]. Computational models of learning like AI2T that model the general processes by which students' knowledge structures change throughout learning have broad benefits toward furthering the science of learning and for analyzing individual learning tasks [20, 35, 56] that go well beyond the affordances of traditional posthoc analyses of student performance data [9].

## 7.2 Broader Authoring with AI2T

One challenge of using AI2T for general-purpose authoring is that it needs to have primitives in its function library to explain and generalize from users' demonstrated actions. Multicolumn addition and fraction arithmetic can be authored with a library that has just a few arithmetic primitives like adding, multiplying, and isolating ones and tens digits from integers. Yet, many domains would require additional pre-built or author-programmed primitives for AI2T to compose into complete programs. There is some precedent for authors learning to write these primitives fairly quickly as part of CMU's LearnLab Summer School. However, in the interest of making these short programming tasks easier, LLM's code generation capabilities could go a long way toward speeding up and helping less-experienced users with writing these primitive functions. Authors' demonstrations could also provide a natural means of unit testing these short sub-routines, and interface affordances could be designed around these features.

Adding yet more performance and learning mechanisms to AI2T could also expand the scope of what it can be used to author with simple primitive functions. For instance, Li et. al. added a representation learning mechanism to SimStudent [33] that greatly simplified the primitive functions it needed to learn algebra equation solving skills. There are similar opportunities along these lines for using pretrained systems like LLMs to parse content like text and images into a structured representation that AI2T can reason over. These features could extend AI2T's authoring scope to include domains with word problems and problems that involve reasoning over figures.

Additionally, some minor improvements in AI2T's process-learning mechanism, like the ability to induce recursive HTNs, would go a long way toward broadening the generality of problems types that can be authored. For instance, in this work we tested 3-by-3 digit multi-column addition, but recursive HTN induction would naturally extend to arbitrary addition problems, and assist in inducing many other domains as well.

### 7.3 Supports for Finding and Fixing Mistakes

Two of our study 2 participants made training mistakes that produced errors that they did not identify and fix, leading to very low final completeness scores. One of the more impactful varieties of mistakes was teaching AI2T an incorrect skill from an incorrect demonstration, or as a result of bad explanations with incorrect formulae or arguments. Since bad skills are relatively easy to identify, eliminating this kind of mistake could be supported by simply adding a feature for removing or editing whole skills, instead of requiring users to delete all of their supporting examples manually.

Mistakes of mislabelling actions' correctness is a harder category of mistake to track down. However, STAND may provide a path toward a solution. Internally STAND filters examples into bins of common examples (like leaves in a decision tree) [57]. It may well be that mistakes and edge-cases tend to filter into bins isolated from the rest. If this is the case, it may be possible for STAND to accurately suggest training examples that it suspects are mistakes.

## 8 Conclusion

Well beyond prior attempts at implementing authoring-by-tutoring [37, 39, 54] this work has demonstrated a path towards methods of ITS authoring that are approachable for non-programmers, yet enable the authoring of flexible and robust ITS behaviors that are typically only implementable with hand-programmed production rules. Prior works with SimStudent and AL have only shown imperfect ITS induction [54, 55] even in the hands of their creators [37, 39, 55]. By contrast, our evaluations of AI2T show half of our untrained participants, who were mostly non-programmers, succeeding at producing model-tracing complete ITSs with authoring-by-tutoring, and several others achieving nearly 100% model-tracing completeness. Further work is needed to adapt AI2T for open-ended authoring and evaluate its efficacy in a wider set of domains. This work presents several interaction design considerations for future authoring-by-tutoring work, including methods for verifying and annotating demonstrations, visualizing and navigating between solution paths, and displaying agents' certainty of proposed actions. STAND [57] and our approach to HTN induction from action sequences mark major machine learning improvements over prior authoring-by-tutoring systems. They enable more data-efficient and robust interactive induction. One of our more surprising results is that STAND's *instance certainty* measure could predict improvements in holdout set performance far better than common ensemble methods like XGBoost, and was used successfully by many of our participants to determine when AI2T had been trained on a sufficient number of problems.